# Science for whom? The influence of the regional academic circuit on gender inequalities in Latin America

Carolina Pradier[1], Diego Kozlowski[1], Natsumi S. Shokida[1], and Vincent Larivière*[1,2,3]

[1] École de bibliothéconomie et des sciences de l'information, Université de Montréal, Montréal, Québec, Canada.

[2] Department of Science and Innovation-National Research Foundation Centre of Excellence in Scientometrics and Science, Technology and Innovation Policy, Stellenbosch University, Stellenbosch, Western Cape, South Africa.

[3] Observatoire des Sciences et des Technologies, Centre interuniversitaire de recherche sur la science et la technologie, Université du Québec à Montréal, Montréal, QC, Canada.

*Corresponding Author

**Email:** carolina.pradier@umontreal.ca; diego.kozlowski@umontreal.ca; natsumi.solange.shokida@umontreal.ca; vincent.lariviere@umontreal.ca

**Abstract**

The Latin-American scientific community has achieved significant progress towards gender parity, with nearly equal representation of women and men scientists. Nevertheless, women continue to be underrepresented in scholarly communication. Throughout the 20th century, Latin America established its academic circuit, focusing on research topics of regional significance. Through an analysis of scientific publications, this article explores the relationship between gender inequalities in science and the integration of Latin-American researchers into the regional and global academic circuits between 1993 and 2022. We find that women are more likely to engage in the regional circuit, while men are more active within the global circuit. This trend is attributed to a thematic alignment between women's research interests and issues specific to Latin America. Furthermore, our results reveal that the mechanisms contributing to gender differences in symbolic capital accumulation vary between circuits. Women's work achieves equal or greater recognition compared to men's within the regional circuit, but generally garners less attention in the global circuit. Our findings suggest that policies aimed at strengthening the regional academic circuit would encourage scientists to address locally relevant topics while simultaneously fostering gender equality in science.

**Introduction**

Scientific production is permeated by inequalities. From the institutional prestige of PhD granting universities (Clauset et al., 2015) to the economic conditions in which researchers conduct their work (Ciocca & Delgado, 2017), scientists carry advantages and disadvantages along their career paths. Individual characteristics that define their identities, such as class (Janke et al., 2017), race (Chen et al., 2022), and gender (Larivière et al., 2013), also affect their career opportunities, as well as the recognition they receive from their work.

At a global level, women are a minoritized group in research—accounting for 29.3% of researchers in 2019 (UNESCO, 2019)—and face greater difficulties in terms of career progression (Diezmann & Grieshaber, 2019; Elsevier, 2020; M. F. Fox, 2020; Huang et al., 2020; Larivière et al., 2013; Shen, 2013; Spoon et al., 2023; Sugimoto & Larivière, 2023). It is widely acknowledged that this situation is detrimental to the development of scientific knowledge, since diverse groups are more likely to make novel scientific contributions (Hofstra et al., 2020) and diversity in the scientific workforce drives the expansion of the knowledge base (Kozlowski et al., 2022, 2024).

Multiple factors shape this phenomenon, ranging from the social roles attributed to men and women—which influence both labor division and topic selection (Bello, 2020; Ceci & Williams, 2011; Shen, 2013)—to issues related to the hostility of the work environment (M. F. Fox, 1991; Spoon et al., 2023), collaboration (Ductor et al., 2023; Holman & Morandin, 2019) and authorship disagreements (Ni et al., 2021). In those parts of the world where parity or almost parity is found, it may not necessarily stem from positive factors: parity may be found in countries where scientific work is not associated with high degrees of social and economic capital, or even signal an unequal brain drain where women have more mobility constraints (Sugimoto & Larivière, 2023).

Latin America is an interesting case study because it is a region[i] where there is a similar number of men and women scientists—according to UNESCO (2019), women account for 45.1% of the scientific workforce—but women are still underrepresented in scientific publishing, and their contributions receive fewer citations (Albornoz et al., 2018; Beigel et al., 2023). They therefore face greater difficulties in getting promoted within the scientific system, even when they exhibit levels of productivity similar to men's (Gallardo, 2022; Jonkers, 2011; López-Bassols et al., 2018). This phenomenon illustrates that women's marginalization in science isn't solely due to their under-representation. It emphasizes the importance of identifying the mechanisms structuring gender inequities in science even in contexts of apparent parity.

While most Latin American countries can be classified as peripheral, Brazil, Mexico, Argentina, Colombia and Chile can be considered semi-peripheral. These countries are not as resource-constrained as the peripheries, but still have access to considerably fewer resources and face more economic and institutional hurdles than the centers (Beigel et al., 2023; Bennett, 2014; Céspedes, 2021). The mechanisms structuring the accumulation of symbolic capital within national scientific fields in the region differ from those of central countries.[ii] Namely, Latin America possesses its own institutions and science policies, along with publication circuits, professional mobility, and consecration processes that function with relative autonomy from the central countries (González et al., 2024). In this context, despite their strong dependency on central countries, one of the most relevant forms of internationalization of the scientific field in these areas is transnational regionalization[iii] (Alatas, 2003; Beigel, 2013, 2014; Heilbron, 2014).

A strong Latin-American academic circuit coexists alongside the global academic circuit, addressing region-specific concerns through articles that are often written in Spanish and Portuguese and disseminated in open access journals, thus ensuring greater circulation and impact within regional boundaries (Beigel, 2014, 2021; Céspedes, 2021; Estrada-Mejía & Forero-Pineda, 2010). The co-existence of regional and global circuits highlights the tension between the two main functions of scientific journals: being a means of communication and an instrument of consecration (Beigel & Salatino, 2015; Salatino, 2018). Latin-American journals usually have a low journal Impact Factor (Paz Enrique et al., 2022), and are therefore considered less prestigious than those published in central countries. In this context, evaluation systems favoring publication in international journals hinder the dissemination of scientific work within the region—due to the paywalls and language barriers researchers must face when papers are published in these journals—and also direct researchers' attention toward topics relevant to central countries (González et al., 2024; van Bellen & Larivière, 2024; Vessuri et al., 2014, p. 650). Despite differences among national evaluation systems, in most countries the symbolic capital accumulated in the international circuit is more easily transferable to the regional circuit than the opposite (Beigel et al., 2023; Beigel & Salatino, 2015; Kreimer, 2011; Ramos Zincke, 2014; Vasen & Lujano Vilchis, 2017; Vélez-Cuartas et al., 2014; Vessuri & Rodriguez-Medina, 2022). However, multiple evaluative cultures and recognition circuits coexist within countries. As a result, international prestige does not directly translate into local prestige in a linear manner (Vessuri & Rodriguez-Medina, 2022).

One of the main challenges faced by the Latin-American scientific system is the insufficiency and instability of its funding, which results from the region's recurrent political and economic crises (Ciocca & Delgado, 2017; Rojas Cifuentes et al., 2023; Salager-Meyer, 2008). This situation has negative effects on scientific production, mainly due to inadequate budgets and salaries (with the resulting difficulties to pay Article Publishing Charges), substandard levels of infrastructure and equipment, and the high cost and limited supply of reagents, among others (Ciocca & Delgado, 2017, p. 847; González et al., 2024; Klebel & Ross-Hellauer, 2022; Salager-Meyer, 2008).

Recurrent crises also influenced the institutionalization and professionalization of science in the region throughout the 20th century. The regional circuit of science was first established in the 1950s along with the emergence of Latin-American intergovernmental institutions that supported the management of scientific information as a key element for development, such as the Economic Commission for Latin America and the Caribbean (ECLAC), the Latin-American Faculty of Social Sciences (FLACSO), and later the Latin-American Council of Social Sciences (CLACSO) (Beigel, 2014; Beigel et al., 2024; Vessuri, 2010). During these years, the number of scientific journals in Latin America started to experience significant growth (Estrada-Mejía & Forero-Pineda, 2010). The circuit was weakened during the 1970s and 1980s with the rise of military dictatorships, and encountered further challenges in the 1990s after democracy was restored as austerity policies brought severe cutbacks on higher education and science. However, it regained strength at the turn of this century with the adoption of policies that once more prioritized the funding of science and higher education (Beigel, 2014), and supported the development of regional systems of indexation and journal quality certification such as SciELO, Latindex, and Redalyc (Beigel, 2013; Beigel et al., 2024). Since the 2010s, however, the resurgence of governments advocating for budgetary constraints has introduced new obstacles to the development of the regional circuit. Specifically, in 2016, for the first time since the year 2000, there was a decrease in the region's Research & Development expenditures (RICYT, 2023). Argentina represents an extreme case, where the national scientific system is currently being dismantled (De Ambrosio & Koop, 2024).

Previous research on the mechanisms structuring gender inequality in science in Latin America has shown the effects of gender stereotypes, unequal distribution of care work and parenting on women's careers (Bello, 2020; Carpes et al., 2022), as well as the presence of strong gender biases in the allocation of science and technology public grants (Fiorentin et al., 2023). Furthermore, productivity gaps between men and women in Argentina and Brazil have been found to widen when considering publication language (Beigel et al., 2023). This study aims to advance our understanding of the integration of Latin-American women and men researchers into global and regional academic circuits. Our goal is to uncover the mechanisms that structure gender inequality in science in Latin America and their relation to the configuration of scientific research topics.

**Materials and Methods**

Data for this article were retrieved from the Dimensions database (Herzog et al., 2020). We examine all publications with at least one Latin America-affiliated author between 1993 and 2022, based on the first institutional affiliation of each author. In this paper, the term "Latin-American researcher" refers to Latin America-affiliated authors. Given the constraints of the available data and the low number of papers for some countries, only the following countries were considered (ordered by the number of publications in the database): Brazil, Mexico, Argentina, Chile, Colombia, Peru, Venezuela, Uruguay, Bolivia, and Paraguay; all of which have more than 2,700 papers in the database over the period studied. Our analysis therefore includes both peripheral and semi-peripheral countries.

Our data consist of 1,845,772 distinct articles and conference proceedings and 3,779,387 total distinct authors, of whom 2,736,608 have a first institutional affiliation in Latin America. Authors were disambiguated using Dimensions' in-house algorithm. While Latin-American authors constitute 72.4% of the distinct authors in the data, they account for 87.64% of all authorships. Authorship is computed using fractional counting, where each publication is divided by its number of authors, and these fractions are then aggregated (the sum of all fractions equals the number of articles in our dataset) to determine the proportion of the articles authored by each group.

The metadata retrieved includes authors' given and family names, which are used to infer gender. The gender disambiguation algorithm builds on the method presented in Larivière et al. (2013) and Sugimoto and Larivière (2023), which uses census data and country-specific lists of men and women names to assign probable gender to researchers based on given names and family names.[iv] Applying this algorithm, it was possible to infer gender for 70.7% of the authors in the dataset, and 72.5% of Latin-American authors. Therefore, our analysis is limited to this subset of the population. Gender is considered in a binary way, as other genders can only be identified through self-identification. This is an acknowledged limitation of the study.

Drawing on the criteria applied by van Bellen & Larivière (2024), we operationalized insertion in the regional academic circuit through publication in a journal located in Latin America (for articles) or publication at a conference held in Latin America (for conference proceedings). This operationalization is informed by the observation that publications in Latin-American journals or conferences tend to have a greater diffusion and impact within regional boundaries. Journal's location information was retrieved from Ulrich's Periodicals Directory using the *ISSN*, which identifies journals present in the Dimensions database; information on the location of conferences was obtained from the Dimensions' database metadata.

Applying the model by Grootendorst (2022), we used articles' abstracts and titles[v] to train a multilingual[vi] semi-supervised[vii] BERTopic model to infer the topic of scientific publications. We considered 100 publications as the minimum topic size, which resulted in a model with 532 topics. As discussed by Rüdiger et al. (2022), it is not advisable to solely rely on automatic evaluation metrics when defining the hyperparameters of unsupervised models because of their dependence on the data and the chosen algorithm. Therefore, we set 100 publications as the minimum topic size based on our analysis of the model's coherence and diversity, but also based on the manual evaluation of the quality of topics identified for each parameter specification, and on the level of detail we sought in our topic space analysis.

The correspondence between topics and disciplines was based on the discipline to which the majority of articles is associated. In cases where no single discipline dominates, the topic is considered multidisciplinary. KeyBERTInspired was used to improve topic representation, which was later refined by hand-labeling to improve their readability. Finally, scholarly impact is assessed through field- and year-normalized citations (Waltman, 2016), using all citations received by articles in our corpus until the end of 2022. In order to examine the audience of different scientific publications, we also compiled the proportion of citations coming from citing articles with at least one Latin-American author. Given the spectrum of possible analysis at the topic level, an extended and interactive version of the results is available at https://vlab.ebsi.umontreal.ca/latam_app/ to allow readers to explore the distribution of all topics.

**Results**

Gendered dissemination circuits

Between 1993 and 2022, Latin-American women have constantly increased their relative participation in scientific publications, both as a proportion of all distinct authors as well as of all authorships—that is, the sum of authors' bylines (Fig. 1A). However, we find a persistent gap between the proportion of women authors and the share of women authorships, as women remain underrepresented in the scientific literature (Fig. 1B). This gap varies over time: it decreased between 1993 and 2009, but has sharply increased since then. As Fig. 1C suggests, women's participation in the regional academic circuit may affect this trend. Between 1993 and 2009, the proportion of scientific documents published in regional venues grew; in turn, the weakened influence of the regional academic circuit since then (Fig. 1C) goes hand in hand with an increase in the productivity gap (Fig. 1B).

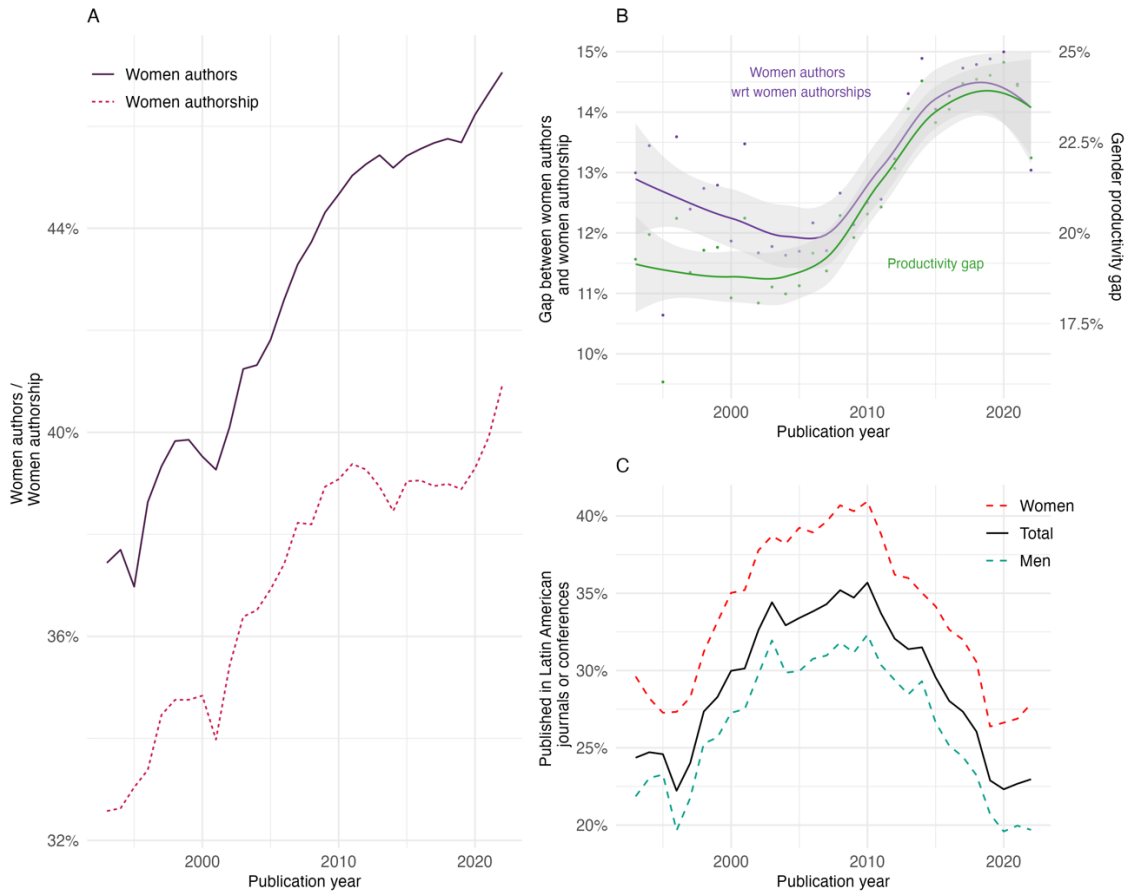

**Figure 1. Women's participation in Latin American science, 1993-2022.** A. Proportion of Latin-American distinct women authors and authorships (fractional count). B. Gap between Latin-American women distinct authors and authorships, and gender productivity gap. LOESS smoothing of the observations (J. Fox & Weisberg, 2019) is used to show the trends in the data with a 95% confidence interval in gray. C. Proportion of documents published by Latin-American researchers that appear in Latin-American journals or conferences.

The number of publications authored by Latin-American researchers grew steadily between 1993 and 2022, with a majority of research published in global venues (Fig. 2). However, growth patterns vary over time and by circuit, discipline, and gender. In the cases of Natural Sciences, Medical and Health Sciences, and Engineering and Technology, the growth in publications within the regional academic circuit stagnated after 2009, while publications in the global academic circuit—particularly publications authored by men—continued to expand. Given the weight of these disciplines in the region's scientific output, this pattern explains the decrease in the relative proportion of regional publications observed in Fig. 1C. However, publications in both circuits grew at comparable rates in the cases Social Sciences, Humanities, and Agricultural Sciences, where both circuits have comparable sizes (Fig. 2). In Social Sciences and Agricultural Sciences, nevertheless, we also observe a sharp increase in the number of publications by men in the global circuit.

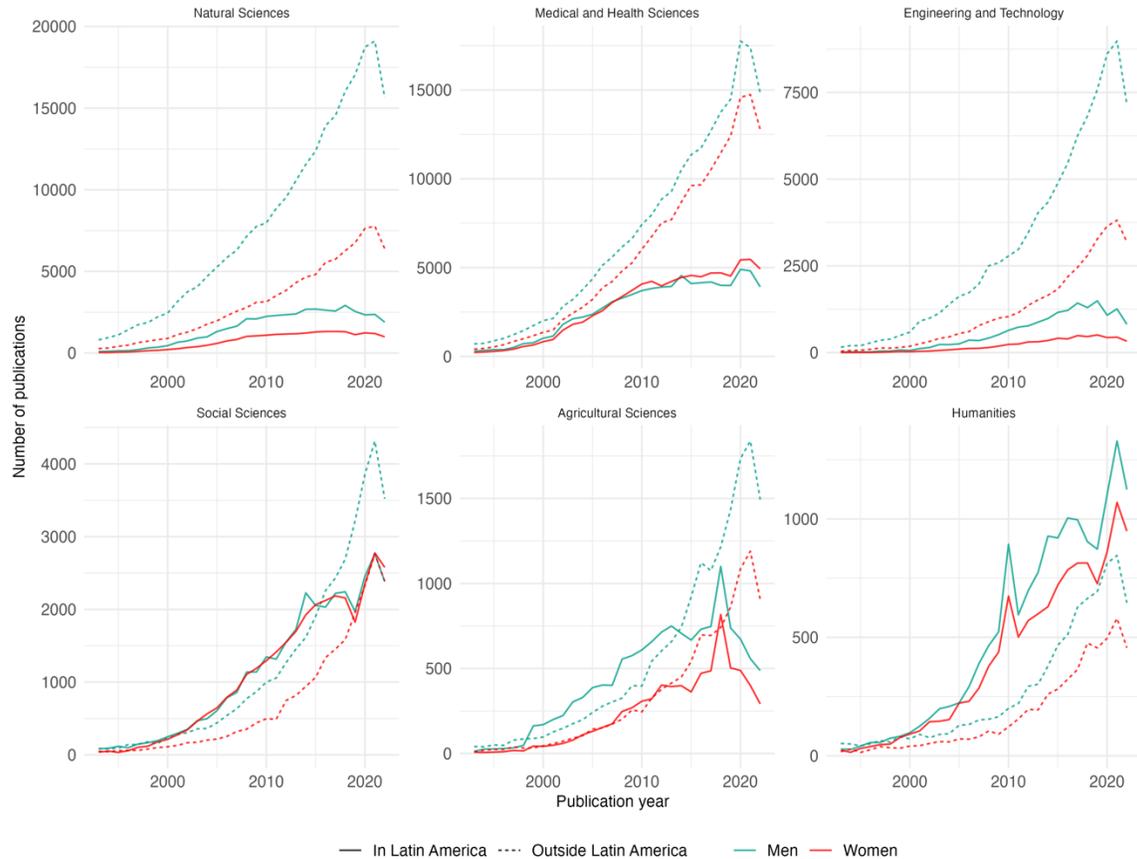

**Figure 2. Publication growth patterns, 1993-2022.** Number of publications by gender circuit and discipline for Latin-American researchers (fractional counts).

To identify the factors contributing to women's greater involvement in the regional academic circuit, we examine its connection to the distribution of men and women across disciplines. Fig. 3A shows the aggregate proportion of documents published in the local circuit by discipline. Following previous work (Vessuri et al., 2014), we observe that Humanities and Social Sciences have greater regional engagement than Natural Sciences and Engineering and Technology, which are almost entirely published in journals from outside Latin America. Finally, Medical and Health Sciences and Agricultural Sciences' substantial contribution to the regional circuit may be related to their high visibility within the SciELO collection (Packer, 2020).

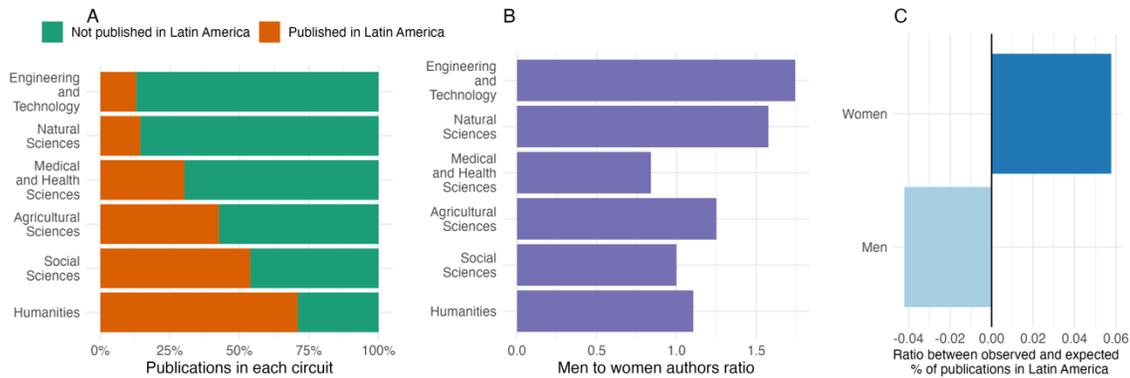

**Figure 3. Disciplines and dissemination circuits by gender, 1993-2022.** A. Proportion of documents authored by Latin-American researchers according to the dissemination circuit and discipline. B. Ratio of men to women Latin-American authors, by discipline. C. Ratio between observed and expected proportion of documents published in Latin-American venues by Latin-American researchers, given the disciplinary distribution of documents from each group.

The degree of regionality of disciplines is partially aligned with the proportion of women in the field. Fig. 3B shows the men to women authors ratio in Latin America by discipline. The two disciplines where men are overrepresented by more than 50%—Engineering and Technology and Natural Sciences—are also those where less than a quarter of the publications are published in regional venues (Fig. 3A). Previous work has shown that Latin America is specialized in Agricultural research (Miao et al., 2022), which can be associated with a relatively higher prestige and internationalization of the discipline in Latin America compared to other regions. The overrepresentation of men in this field (Fig. 3B) is coherent with those findings. Fig. 3A and 3B suggest that a relevant factor for the overrepresentation of women in the local circuit is the gendered distribution across disciplines, where women are more engaged in disciplines that prominently circulate within the local circuit. However, this is not the only factor in play. By computing the expected proportion of publications in local journals as a function of the distribution by discipline, Fig. 3C controls for the disciplinary effect. Using such normalization, we find that women systematically publish more than expected within the regional academic circuit (Fig. 3C).

Publication venues and topic space

The differences in women's and men's involvement in the global and regional academic circuits seems to be a fundamental factor driving their unequal representation in scientific publishing. However, our results show that these disparities cannot be solely attributed to differences in the gender distribution of researchers by discipline. More specifically, women and men within the same discipline are likely to disseminate their work through different circuits. Given previous research suggesting that the Latin-American scientific community is primarily organized around research themes rather than disciplines (González et al., 2024), we extracted the topic of publications (see Materials and Methods). Fig. 4 illustrates the proportion of publications within each topic disseminated through the regional circuit (y-axis), while simultaneously considering the degree of feminization of each topic (x-axis).

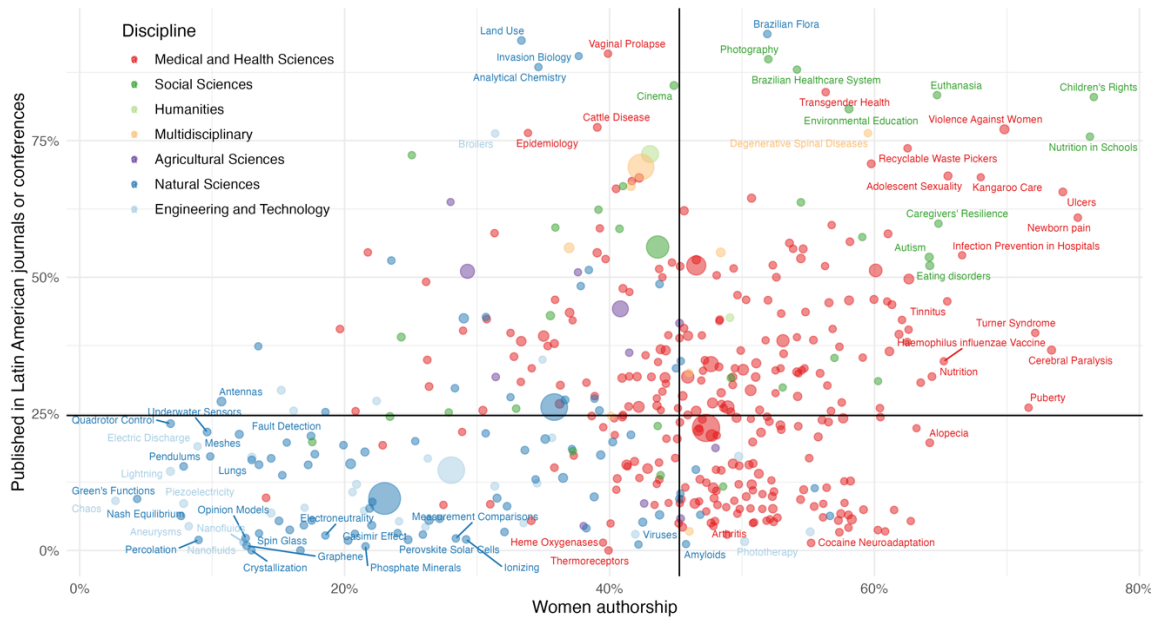

**Figure 4. Distribution of topics by dissemination circuit (y-axis) and degree of feminization (x-axis).** The horizontal axis shows the proportion of women authorship within each topic, while the vertical axis shows the proportion of articles and conference proceedings published in Latin America. Size represents the absolute number of publications in the topic. The figure displays 453 topics, as topics with less than 50 documents with complete information on feminization and dissemination circuit were excluded. Topics with the 20 highest and lowest rates of women authorship and the 20 highest and lowest proportions of publications in Latin America were hand-labelled.

Results show that most of the topics in Natural Sciences and Engineering and Technology are in the bottom-left quadrant, indicating both a low proportion of women authorship and a low proportion of publications in the regional circuit. Notable exceptions are topics related with *Land Use, Brazilian Flora*, and *Invasion Biology*, which are highly regional, escaping their fields' logic to engage mostly with the global circuits of dissemination. Meanwhile, Medical and Health Sciences cover topics which circulate on a broader range of circuits, and therefore span across the four quadrants, with an overall higher degree of feminization. Social Sciences and Humanities' topics are predominantly located in the top-right quadrant, presenting a high proportion of publications in the regional circuit while also displaying high levels of feminization.

A more detailed analysis of the most feminized topics reflects women's focus on children and adolescents' well-being (such as *Children's Rights, Newborn Pain, Kangaroo Care,* and *Adolescent Sexuality*), while considering issues specifically relevant to the region such as the high prevalence of eating disorders (Kolar et al., 2016). Moreover, they address issues particularly relevant to low-income countries, such as *Infection Prevention in Hospitals* or *Nutrition in Schools*. Among the topics discussed within the regional circuit, we find some of the main challenges the region currently faces to ensure its population's wellbeing, such as *Recyclable Waste Pickers*' health (Binion & Gutberlet, 2012; Kain et al., 2022), *Transgender Health* (Campbell, 2019; Corrales, 2015; Socías et al., 2014), *Violence Against Women* (Batthyány, 2023; Guedes et al., 2014; Velásquez et al., 2020) and *Land Use* (Angotti, 2013; Guereña & Burgos, 2016; Stocks, 2005)*.*

Those results illustrate the configuration of each discipline's topic space across diffusion circuits and degrees of feminization, and show—in all disciplines—a clear overlap between women's research interests and issues specifically relevant to Latin America. As suggested by previous literature studying the relationship between publication venue and research topic (Estrada-Mejía & Forero-Pineda, 2010), we can conclude that women's focus on topics that are directly relevant for their communities is one of the main factors that lead them to disseminate their findings through local journals or conferences.

<u>Scholarly impact and dissemination circuits</u>

To better understand how topic choice affects career paths, we focus on how scientific knowledge circulates in Latin America, as well as how symbolic capital is accumulated. For this, we divided topics into two categories, based on the proportion of publications in each dissemination circuit: 1) Global topics, with a proportion of publications in Latin-American venues that is below average (23.93%), and 2) Regional topics, where that proportion is above the average. We then compile the distribution of averaged normalized citations (Fig. 5A) and proportion of citations received coming from Latin America (Fig. 5B).

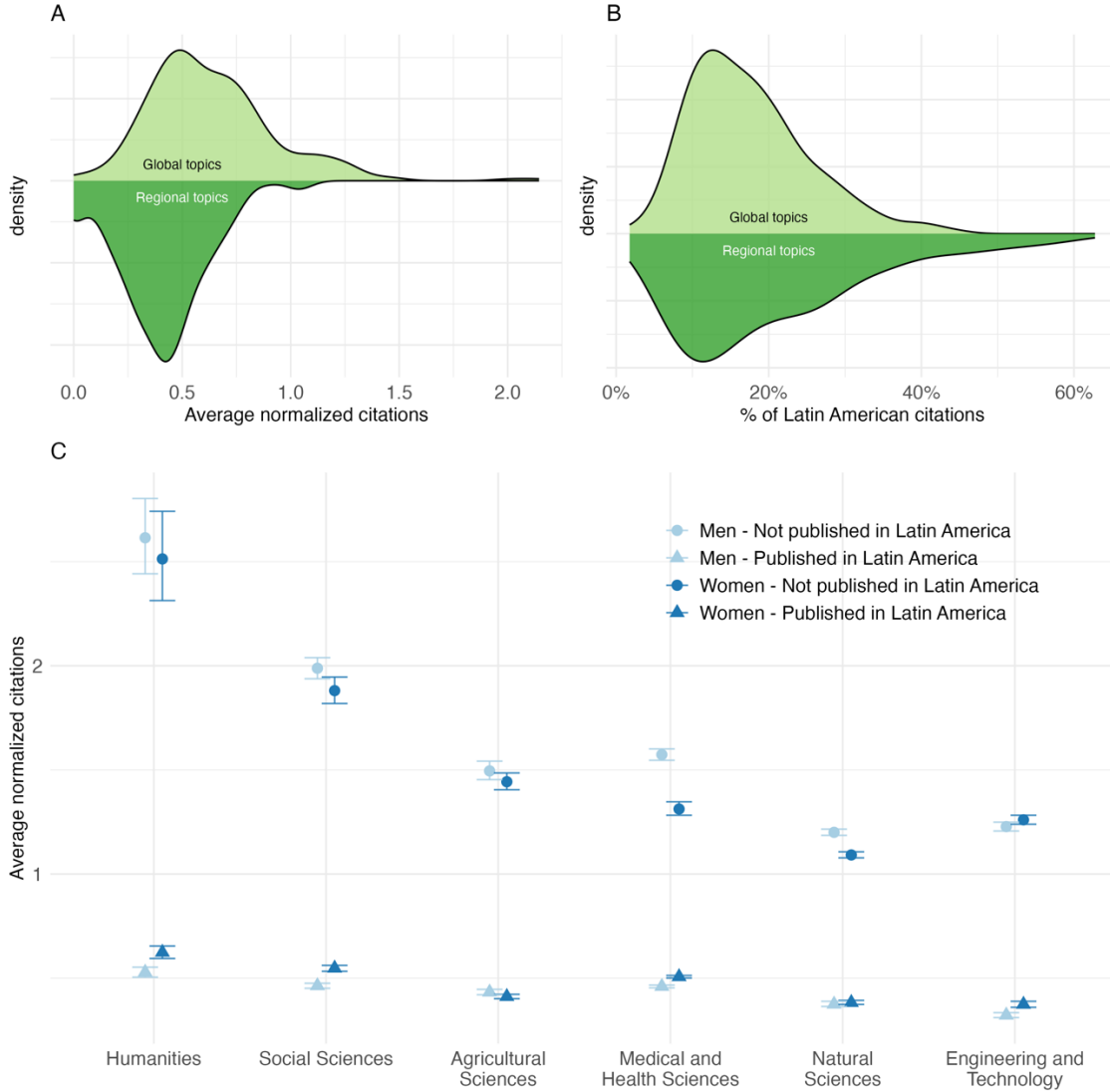

**Figure 5. Scholarly impact, research topics and publication venue.** A. Distribution of field-normalized citations for each topic. B. Distribution of the proportion of citations from articles with at least one Latin-American author for each topic. In Figures A and B, topics are classified as global if the proportion of publications from Latin America within each topic is below the average (23.93%), and as local if it is above the average. C. Average field-normalized citations of publications by discipline, gender and dissemination circuit with a 95% confidence interval resulting from 1000 bootstrap resamples.

Unsurprisingly, publications on regional topics receive fewer citations on average (Fig. 5A). Citations received by these papers are more likely to come from other Latin-American researchers (Fig. 5B). These findings suggest that scientists working on those topics will be at a disadvantage in citation-based evaluations compared to those who focus on topics of global interest. As discussed by Velho (2005), these lower citation rates do not imply that research is of lower quality. Citations are influenced by several social factors, such as higher credibility attributed to research from central countries, as well as the size of the research communities and national self-citations (Baccini et al., 2019; Baccini & Petrovich, 2023; Gomez et al., 2022). Moreover, such lower citation rates are at odds with the intrinsic value of this work, as this research is of strategic relevance for the scientific development of the region. Our findings highlight the central role of the regional academic circuit in the dissemination of research that addresses issues specifically relevant to the region (Fig. 4) even if it receives less global attention.

While there is a general trend to gain more recognition by publishing outside the region, the mechanisms structuring symbolic capital accumulation can be further examined in terms of their relation to discipline and gender. Fig. 5C combines these dimensions by showing the average normalized citations by discipline, gender, and dissemination circuit. Even though there is a significant citation gap favoring the global circuit over the regional circuit in all cases, the citation gap between articles published outside Latin America and those published within the region varies across disciplines and also differs by gender (Fig. 5C). More specifically, the citation gap between circuits is more pronounced in Humanities and Social Sciences, the disciplines where most of the work is published within the local circuit. Nevertheless, Agricultural Sciences, Medical and Health Sciences, Natural Sciences and Engineering and Technology all present significant citation gaps between circuits, with Natural Sciences displaying the smallest gap.

Fig. 5C also shows that research authored by men receives more citations than that by women when published outside the region, which is coherent with the general finding—based on Western databases such as the Web of Science—that women's work is less cited on average (Sugimoto & Larivière, 2023). However, such finding is not observed in Engineering and Technology, where women's work published outside the region is slightly more cited, on average, than that of men. This result echoes previous work, which has shown that at the world level, women in Engineering tend to publish their work in journals with higher Impact Factors than men (Ghiasi et al., 2015), and where the gender gap in citations is much smaller than in the medical field (Sugimoto & Larivière, 2023). Within the regional academic circuit, however, women's work is either equally or more cited than men's work, and this finding is observed in all disciplines. Overall, these results show the configuration of an elite, where for those fields where the regional circuit is the norm, international publishing acts as a symbolic marker of distinction—which results in a huge difference in recognition—, while for fields where most of the work is published internationally, the reward for disseminating articles through the global instead of the regional circuit is smaller.

In summary, these results suggest that gender differences in scholarly impact result from the interplay between differences in discipline and topic choice, and the existence of distinct mechanisms guiding symbolic recognition within each circuit. Due to their topic and discipline choice, Latin-American women are underrepresented in the global academic circuit, where publications systematically receive more citations. Additionally, comparing citations received by men and women's publications within each circuit, we find that women's work receives equal or more recognition than men's publications when published in the regional circuit, but systematically receives less attention when disseminated in the global circuit.

**Discussion**

Gender inequalities in science are shaped by the social context in which science is produced. To understand how gender inequalities are shaped in Latin America, it is therefore essential to consider the role of the region's academic circuit. Our findings show a consistent increase in women's involvement in science over the last 30 years, as well as a higher participation of women in the academic regional circuit compared to men—irrespective of their distribution across disciplines. Our results also suggest that, although the region would greatly benefit from strengthening its regional academic circuit, which is used by researchers to disseminate findings that are regionally relevant, political and economic instability have historically affected its development. Future research should further examine other relevant elements shaping the Latin American scientific output, such as open access practices, the role of foreign funding, and national differences.

Our findings are consistent with the coexistence of two differentiated circuits of knowledge diffusion. Those circuits structure gender inequality in Latin America in terms of representation and impact. Research disseminated through the regional circuit addresses local audiences and is considerably less cited than research published outside the region. This implies that a fundamental factor structuring gender inequality in the distribution of recognition within this region is the greater involvement of women in the regional academic circuit, which focuses on the region's specific needs. In this regard, we also found that the mechanisms structuring gender differences in symbolic capital accumulation in each circuit differ: while women's work receives equal or more recognition than men's when published in the regional circuit, it systematically receives fewer citations when it is disseminated in the global circuit.

In this context, the lower relative participation of women in science in Latin America not only undermines the region's scientific production due to the increased propensity of underrepresented groups to produce innovative outcomes (Hofstra et al., 2020), but also in terms of women's propensity to tackle issues of regional relevance. In this regard, evaluation systems favoring publication in international journals orient researchers' focus towards topics relevant to central countries (Vessuri et al., 2014). In summary, we argue that the Latin-American scientific system faces two major challenges: first, globally fostering greater gender diversity across topics and disciplines; and second, strengthening the regional academic circuit to ensure equitable recognition for researchers addressing locally relevant topics.

**Acknowledgments**

This project was funded by the Social Science and Humanities Research Council of Canada Pan-Canadian Knowledge Access Initiative Grant (Grant 1007-2023-0001), and the Fonds de recherche du Québec - Société et Culture through the Programme d'appui aux Chaires UNESCO (Grant 338828). The authors wish to thank Lucía Céspedes for her comments on the manuscript, and Thema Monroe-White, Monica Novoa, and Cassidy R.

---

[i] Throughout the article, we will use the term 'region' to denote Latin America and the Caribbean, as per the United Nations Regional Groups classification.

[ii] In this context, we expand on Alatas' (2003) conceptualization to define central countries to be those that command considerable recognition and prestige and can therefore determine research agendas, problems areas, methods of research, and standards of excellence. Additionally, scientists in these countries tend to consider the disclosure of the location of their studies to be largely irrelevant due to the universal significance they attribute to their findings (Baber, 2003; Castro Torres & Alburez-Gutierrez, 2022).

[iii] Transnational regionalization can be defined as "a spatially integrated form of political co-operation and problem-solving that transcends the limits of nationally based administrative practice and attempts to create a sense of cohesiveness, interdependence and common interests across national boundaries" (Scott, 2002, p. 179).

[iv] In the case of certain countries such as Ukraine, family names are necessary to infer gender (note that Latin-American author's collaborators come from all over the world).

[v] Drawing on the criteria applied by Kozlowski et al. (2021), for each article we built a text that contains three times the title of the publication and once the abstract.

[vi] A multilingual model was employed because articles in the corpus were written in English, Portuguese and Spanish (along with a small proportion of documents in other languages). Stopwords in these three languages were removed, as well as words belonging to snippets of LaTeX code that could be found in the data.

[vii] Using discipline labels provided by Dimensions.